# Deep learning for the design of non-Hermitian topolectrical circuits

Xi Chen, [1, §] Jinyang Sun, [2, §] Xiumei Wang, [3] Hengxuan Jiang, [1] Dandan Zhu, [4†] and Xingping Zhou [5‡]

[1] *College of Integrated Circuit Science and Engineering, Nanjing University of Posts and Telecommunications, Nanjing 210003, China*
[2] *Portland Institute, Nanjing University of Posts and Telecommunications, Nanjing 210003, China*
[3] *College of Electronic and Optical Engineering, Nanjing University of Posts and Telecommunications, Nanjing 210003, China*
[4] *Institute of AI Education, Shanghai, East China Normal University, Shanghai 200333, China*
[5] *Institute of Quantum Information and Technology, and Key Lab of Broadband Wireless Communication and Sensor Network Technology, Ministry of Education, Nanjing University of Posts and Telecommunications, Nanjing 210003, China*
[†] *ddzhu@mail.ecnu.edu.cn*
[‡] *zxp@njupt.edu.cn*
§ *These authors contributed equally to this work.*

Non-Hermitian topological phases can produce some remarkable properties, compared with their Hermitian counterpart, such as the breakdown of conventional bulk-boundary correspondence and the non-Hermitian topological edge mode. Here, we introduce several algorithms with multi-layer perceptron (MLP), and convolutional neural network (CNN) in the field of deep learning, to predict the winding of eigenvalues non-Hermitian Hamiltonians. Subsequently, we use the smallest module of the periodic circuit as one unit to construct high-dimensional circuit data features. Further, we use the Dense Convolutional Network (DenseNet), a type of convolutional neural network that utilizes dense connections between layers to design a non-Hermitian topolectrical Chern circuit, as the DenseNet algorithm is more suitable for processing high-dimensional data. Our results demonstrate the effectiveness of the deep learning network in capturing the global topological characteristics of a non-Hermitian system based on training data.

## I. INTRODUCTION

Deep learning is a class of machine learning algorithms that is primarily focused on image processing [1]. It uses multiple layers to describe high-level features from the raw input. For example, in image processing, low-level features, such as color, texture, orientation, edge, etc., while higher layers may identify the concepts relevant to humans such as faces or traffic conditions. Recently, multimodal learning [2] and continual learning [3] have gained prominence as methods to address complex data processing tasks. These algorithms and modeling tools show extremely good results especially for large models such as chat Generative Pre-trained Transformer (chat-GPT) [4].

On the other hand, exotic topological states and phenomena in non-Hermitian systems has seen an explosion of activity via different physical platforms, such as photonics, mechanical systems, and electric circuits [5]. The non-Hermiticity may come from gain and loss effects [6-9], nonreciprocal hoppings [10-12], or dissipations in open systems [13], leading to the breakdown of conventional bulk-boundary correspondence [12, 14, 15] and the non-Hermitian topological edge mode [9,16]. Although the non-Bloch bulk-boundary correspondence and non-Bloch Chern numbers have been generalized in Ref [12] and Ref [14] to solve breakdown of the bulk-boundary correspondence, the extreme sensitivity of the spectrum to the boundary condition in non-Hermitian system has still attracted tremendous attention [17-23].

Deep learning as a powerful tool for data analysis has been introduced into a number of physical settings research recently due to its ability in approximating any continuous functions, ranging from black hole detection [24], gravitational lenses [25], photonic structures design [26], quantum many-body physics, quantum computing, and chemical and material physics [27]. In particular, the research of topological invariants in Hermitian systems has been actively pursued in the last few years based on the deep learning approach with various algorithms and models [28-34]. The topological phases in disordered systems are also investigated [35]. Actually, deep learning is more suitable for the non-Hermitian system due to the complexity of the topological characteristic. The neural networks are employed to identify exceptional points (EPs) in non-Hermitian systems, predicting EPs in various models [36]. Some previous work in non-Hermitian systems based on deep learning successfully show the ability to predict the topological phases [37-41]. The inverse-design for non-Hermitian systems can also be realized via machine learning [42]. But they mainly focus on the prediction with one or two feature inputs. We will present the ability of deep learning algorithms in handling with five-dimensional inputs.

In our work, we present the application of deep learning to predict non-Hermitian topological phases and design a non-Hermitian Chern circuit. Firstly, we investigate the predictive performance of Multilayer Perceptron (MLP) [43] and Convolutional Neural Network (CNN) [44] models on the winding number of a one-dimensional non-Hermitian Su-Schrieffer-Heeger (SSH) model [12]. The results demonstrate that both models exhibit similar levels of train loss and test accuracy, with accuracies surpassing 98%. We attribute the minimal differences between the two networks to the low-dimensional nature of the input data and the absence of noise. Our findings indicate that MLP and CNN networks are capable of accurately predicting winding numbers, suggesting their suitability for this task and highlighting the potential of deep learning in identifying non-Hermitian topological phases based on winding numbers.

Then, we explore the manifestation of Chern numbers in a globally periodic circuit following the structure in Ref [45], which serves as an extension on winding numbers. We investigate the realization of Chern states with arbitrary Chern numbers in a circuit. Specifically, we concentrate on the cases of $C=1$ and $C=0$, where the distinctions between these two different Chern numbers are reflected in the voltage distributions across the circuit units. For $C=0$, the voltage exhibits a disordered random distribution, whereas for $C=1$, the voltage distribution displays edge states. Our approach involves extracting high-dimensional circuit data features from the minimum module of the periodic circuit and utilizing Dense Convolutional Network (DenseNet) [46], a convolutional neural network, to design a non-Hermitian topolectrical Chern circuit. Our results demonstrate that deep learning networks can capture global topological features in non-Hermitian systems from training data. Furthermore, to explore the reliability and generality of the model, we randomly select a set of circuit data and remove a 2×2 unit block from the bottom of the circuit. The results indicate that our network exhibits high generality and stability. It achieves an accuracy of approximately 99.5% in predicting the state of the complete circuit and accurately predicts the state values of defective circuits.

## II. TOPOLOGICAL INVARIANTS PREDICTION IN SU-SCHRIEFFER-HEEGER MODEL

The non-Bloch band theory was initially proposed to solve the problem of body edge correspondence for non-Hermitian systems under open boundary conditions (OBC) [12]. In Hermitian systems, there is a correspondence between the topological invariant contained in Bloch Hamiltonian under periodic boundary conditions (PBC) and the topological protected boundary states under OBC. However, in some non-Hermitian systems, the energy spectrum of Bloch Hamiltonian and its corresponding wave function in the form of modulated plane wave are significantly different from the energy spectrum and wave function under OBC, which means that the topological properties of non-Hermitian Bloch Hamiltonian cannot predict the behavior of boundary states under OBC [47].

Due to the limitations of the Bloch Hamiltonian to account for the non-Hermitian skin effect in open boundary conditions, the Bloch topological invariant may not provide an accurate prediction of the topological properties of non-Hermitian systems under such conditions. At present, the non-Bloch energy band theory can address this issue by incorporating the non-Hermitian skin effect [12, 14, 15]. In the non-Bloch energy band theory, the topological invariant is no longer defined in the traditional Brillouin zone, but in the generalized Brillouin zone. This topological invariant describes the topological properties implied by the Hamiltonian on the generalized Brillouin zone, so it is called a non-Bloch topological invariant.

We propose to begin our analysis with the SSH model following Ref [12], which is a typical non-Hermitian model that exhibits topological phases. The non-Hermitian Su-Schrieffer-Heeger (SSH) model, whose Bloch Hamiltonian is

$$H(k) = d_x \sigma_x + (d_y + i\frac{\gamma}{2})\sigma_y, \tag{1}$$

where $d_x = t_1 + (t_2 + t_3)\cos k$, $d_y = (t_2 - t_3)\sin k$, and $\sigma_{x,y}$ are the Pauli matrices, $t_1$ and $t_2$ represent transitions within and between protocells, respectively, and $\gamma$ represents non-Hermitian terms.

If the Bloch Hamiltonian is extended to the $\beta$ complex plane, the Hamiltonian of the non-Hermitian SSH model can be written as:

$$H(\beta) = (t_1 - \frac{\gamma}{2} + \beta t_2)\sigma_- + (t_1 + \frac{\gamma}{2} + \beta^{-1} t_2)\sigma_+, \tag{2}$$

where $\sigma_\pm = (\sigma_x \pm i\sigma_y)/2$, $\beta = re^{ik}$. Here, $r = \sqrt{|(t_1 - \gamma/2)/(t_1 + \gamma/2)|}$ is the radius of the generalized Brillouin zone, and k is a real parameter between $[0, 2\pi]$. The chiral symmetry of the non-Hermitian SSH model makes $|\tilde{u}_{R,L}\rangle = \sigma_z |u_{R,L}\rangle$ the right and left vectors of energy $-E(\beta)$, which satisfy the following normalization relation: $\langle u_L | u_R \rangle = \langle \tilde{u}_L | \tilde{u}_R \rangle = 1$, $\langle u_L | \tilde{u}_R \rangle = \langle \tilde{u}_L | u_R \rangle = 0$. From this, the Q matrix can be defined:

$$Q(\beta) = |\tilde{u}_R(\beta)\rangle\langle \tilde{u}_L(\beta)| - |u_R(\beta)\rangle\langle u_L(\beta)|, \tag{3}$$

where $|u_{R,L}\rangle$ represent separately the right and left arrows of $H(\beta)$. The chiral symmetry $\sigma_z Q \sigma_z = -Q$ makes it have the following anti angular form:

$$Q = \begin{pmatrix} 0 & q \\ q^{-1} & 0 \end{pmatrix}. \tag{4}$$

At this time, the non-Bloch linking number can be defined as the integral along the generalized Brillouin zone:

$$W = \frac{i}{2\pi} \int_{GBZ} q^{-1} dq. \tag{5}$$

For the SSH model, the number of cycles in the non-Hermitian topological phase is $W=1$, while the trivial phase is $W=0$. Then we consider the topological phase prediction in deep learning method with one and two feature inputs.

(1) One feature input

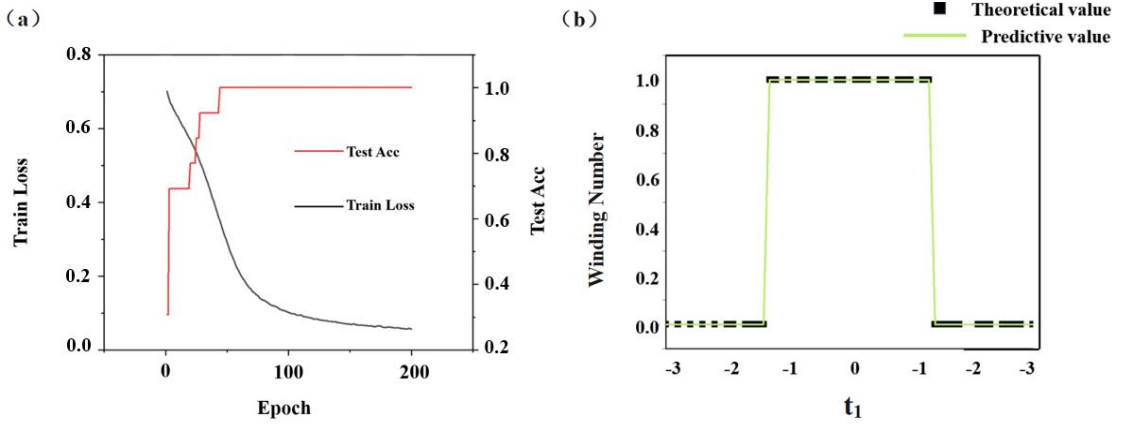

FIG. 1. A fully connected neural network with one-dimensional feature inputs for the SSH model. A fully connected neural network with two hidden layers is constructed. The hidden layer consists of 100 and 32 neurons, respectively. The training set consists of 61 samples. (a) Train loss and test acc for each training epoch of the MLP. (b) Theoretical and predicted values of winding numbers of the deviation value at $t_1 \in [-3,3]$, $t_2 = 1$, $\gamma = 1$.

Our analysis starts by establishing the SSH model and feeding a single feature into the network. Next, a fully connected neural network with two hidden layers is constructed. The first hidden layer consists of 100 neurons and the second hidden layer consists of 32 neurons. The hidden layer adopts the rectified linear unit (ReLU) activation function. In the output layer, we use the Sigmoid activation function. In order to train our neural network, a training set consisting of 61 samples is generated. We set step size of $t_1$ to 0.1, to generate a training set. The network is trained for 200 epochs with a batch size of 50. The loss of each training stage is shown in Fig. 1 (a). We note that the neural network converges quickly. It is because the neural network converges quickly that we check the precision of each training cycle of the training set and the test set. Besides, we find that their precision is approximately equal, which rules out overfitting. After training the neural network on the training set, we can utilize the network to make predictions on the test set, which consists of data that the network has not encountered during the training phase. The test set has a size of 20%, which is not seen during training. The predicted number of turns $W$ is shown in Fig. 1 (b). We observe that our trained neural networks produce winding numbers that are very close to integer values. As is common practice, we round the output to the nearest integer. It is found that the neural network we trained shows a very high accuracy, exceeding 99.9%. As can be seen from the above, MLP exhibits a good degree of fitting for the input $t_1$. Next, we will further explore the network's requirements and identify the optimal fit for input data with two features, $t_1$ and $t_2$.

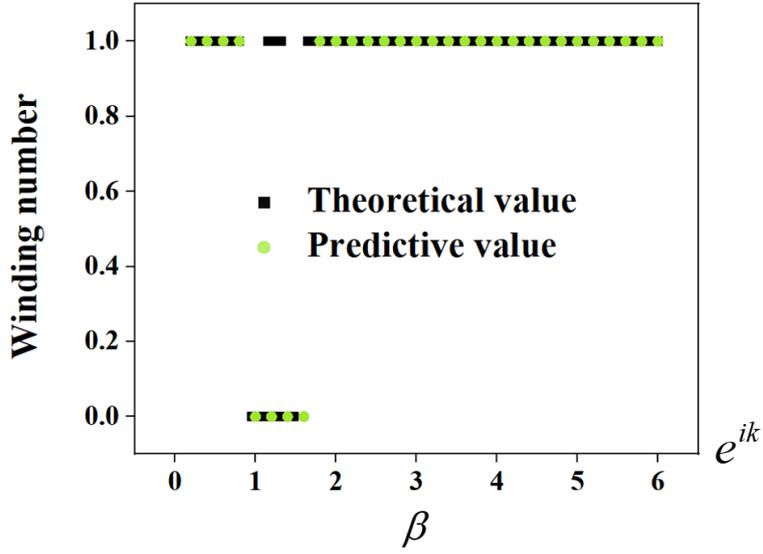

FIG. 2. Theoretical and predicted values of winding numbers of the deviation value at $\beta \in [0,6]$.

At the same time, we draw the relationship between the root $\beta$ (Eq. 2) and the winding number $W$, as shown in FIG. 2, where W indicates the presence of skin effect and bulk localization. The results show that the accuracy of the network is still high.

By definition, the skin effect is determined by whether the topological invariant is zero. When the topological invariant (i.e. the winding number) is non-zero, the skin effect appears; but when the winding number is equal to zero, there is no skin effect. Therefore, the key to determine whether the system exhibits skin effect is to assess if the winding number is non-zero. According to Eq. (2), the Hamiltonian formula of the non-Hermitian SSH model includes three variables $t_1$, $t_2$ and $\gamma$. In order to reduce the complexity, we set $t_1 \in [-3,3]$, $t_2 = 1$, $\gamma = 1$. The function of $\beta$ is to extend the Hamiltonian to the complex plane, and $\beta = re^{ik}$, $r = \sqrt{|(t_1 - \gamma/2)/(t_1 + \gamma/2)|}$, which shows that the $\beta$ is also calculated from the $t_1$ variable, just like H($\beta$). To sum up, the variable $t_1$ determines H($\beta$), and H($\beta$) determines the winding number of zero or non-zero. The $\beta$ also contains the variable $t_1$. Therefore, it exists $\beta \in$ (0,1) ∪ (1.5,6) in Fig. 2 makes the winding number $W$ non-zero or zero when $t_1 \in [-3,3]$ changes, which has a skin effect or bulk localization.

(2) Two feature inputs:

For the two feature inputs, we adapt deep learning with MLP and CNN architectures to predict non-Hermitian topological invariants.

1. Multi-Layer Perceptron (MLP):

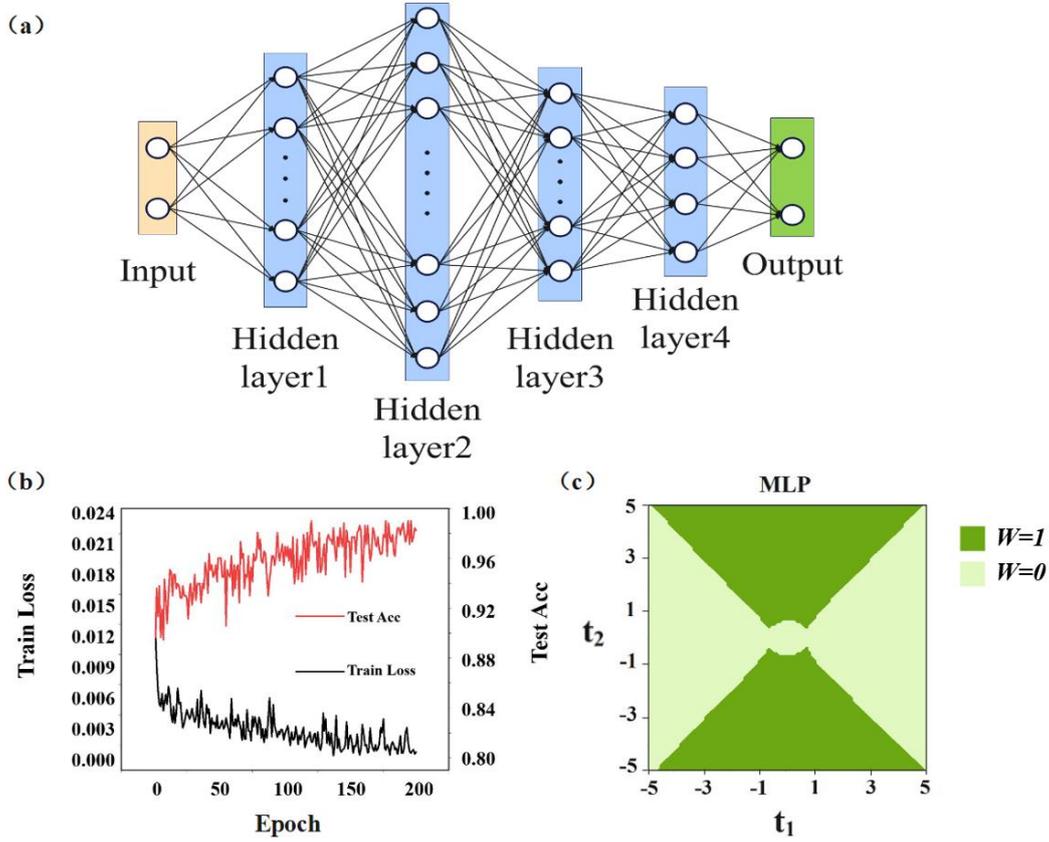

FIG. 3. Fully connected neural network for 2D feature input of SSH model. (a) Schematic diagram of our MLP network with four hidden layers, which are 12, 24, 8, and 4 neurons respectively, and then two neurons are used in the output layer to predict "0" and "1" results. (b) The train loss and test accuracy (acc) of each training epoch. (c) The prediction of winding numbers by the MLP network, where $t_1 \in [-5,5]$, $t_2 \in [-5,5]$, $\gamma = 1$.

Similar to the MLP network with a single feature input, we continue to use a fully connected neural network in this case. The input layer of the neural network accepts input from 2 features. The schematic diagram of our MLP neural network is shown in Fig. 3(a). For the output layer (single neuron), the Sigmoid activation function is used to generate the prediction results of binary classification. It can be seen that the accuracy of this network can maintain around 98% after sufficient training shown in Fig. 3(b), while the prediction of winding numbers is shown in Fig. 3(c).

2. Convolutional Neural Network (CNN):

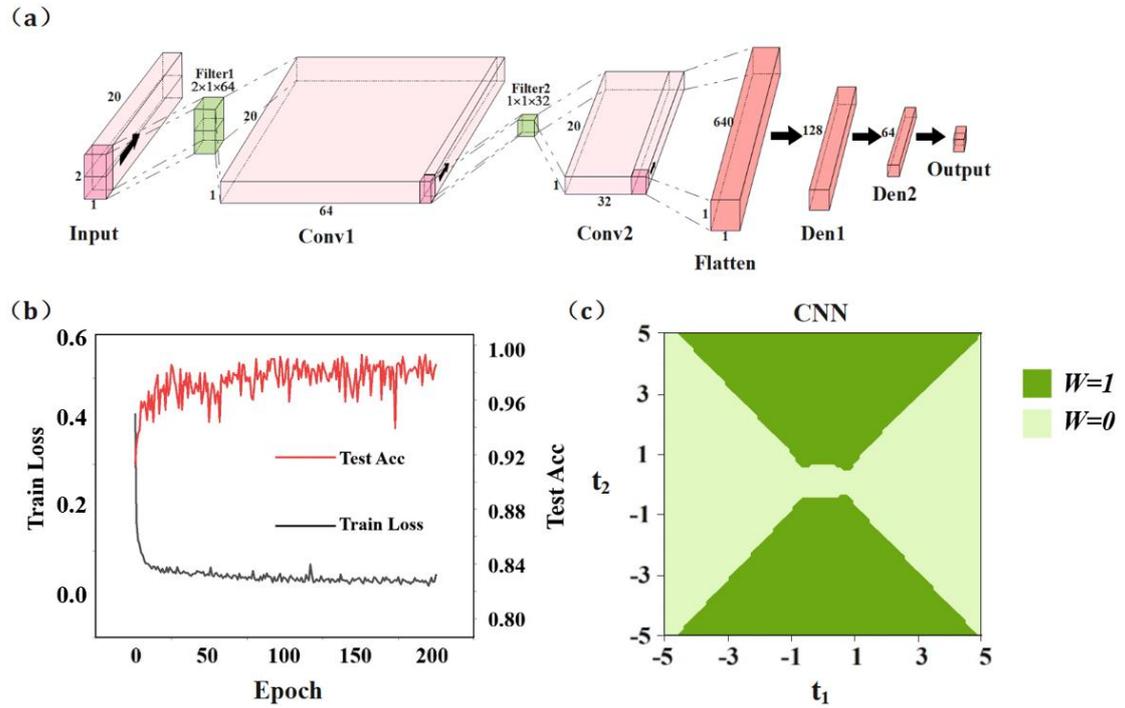

FIG. 4. Convolutional neural network for 2D feature input of the SSH model. A convolutional neural network with a Conv1D layer, a Flatten layer, and two Dense layers (activation functions ReLU and Sigmoid) is constructed. The training set consists of 3721 samples. (a) Schematic of our convolutional neural network with two convolutional layers, with 64 and 32 filters and kernel size of $2 \times 1$ and $1 \times 1$, followed by a flatten layer and two dense layers before the output layer, which are used to predict the winding numbers. (b) The train loss and test acc of each training epoch. (c) The prediction of winding numbers by the CNN network, where $t_1 \in [-5,5]$, $t_2 \in [-5,5]$, $\gamma = 1$.

Due to the increase in the input dimensionality and the growth in the number of columns in the dataset, we choose to proceed with training using a CNN, which allows for batch processing of data during training. The network structure consists of two Conv1D layers, a Flatten layer, two Dense layers (with ReLU activation), and a final output layer (with Sigmoid activation), which are designed to improve the performance of our classification task. We present a schematic of our CNN architecture, consisting of two convolutional layers. The first convolutional layer consists of 64 filters with a kernel size of 2×1, while the second convolutional layer comprises 32 filters with a kernel size of 1×1. Subsequently, a flatten layer is applied to transform the output into a one-dimensional representation. Following the convolutional layers, we add two fully connected layers. The first fully connected layer utilizes the ReLU activation function to compress the input to 128 neurons. Similarly, the second fully connected layer also employs the ReLU activation function, further compressing the input to 64 neurons. The schematic of our proposed network structure is shown in Fig. 4(a). To facilitate prediction, we conclude the network architecture with an output layer employing the Sigmoid activation function. This output layer consists of two neurons, enabling the prediction of the number of convolutions as either "0" or "1". It can be seen that the accuracy of the network can be maintained at about 98% after enough training as shown in Fig. 4(b) and the prediction of winding numbers is shown in Fig. 4(c).

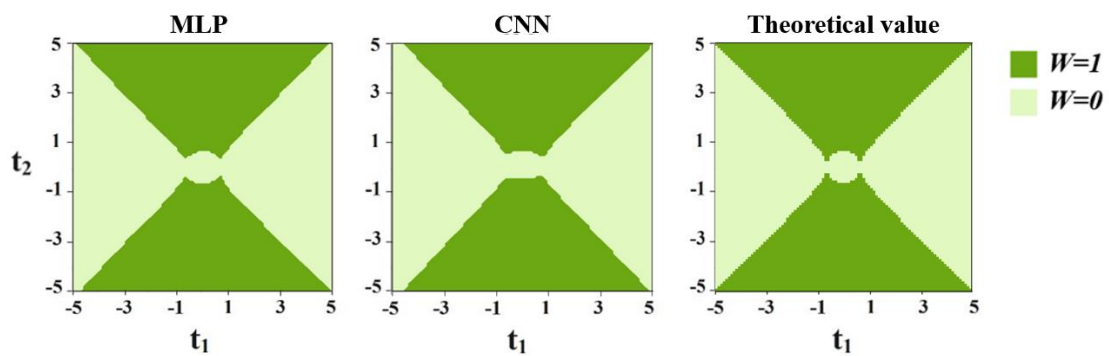

FIG. 5. The comparison between prediction winding values of MLP, CNN and the theoretical winding values, where $t_1 \in [-5,5]$, $t_2 \in [-5,5]$, $\gamma=1$. The accuracy of both MLP and CNN have reached a high level.

As shown in Fig. 5, although the accuracy of the winding number predicted by the MLP network is seen better than that of CNN, the accuracy of both can reach more than 98%. We attribute the slightly worse effect of CNN than MLP to the low dimension of the input data. In fact, both have achieved good results, which shows that neural network has broad prospects in predicting topological invariants.

In this section, we conduct a performance comparison between MLP and CNN architectures for predicting the winding number using 1D and 2D data inputs. Despite their architectural differences, both algorithms exhibit comparable accuracy and loss metrics. However, the CNN algorithm demonstrates notably faster computational speed compared to MLP when operating on equivalent data volumes, which provides a solution for us to quickly process high-dimensional datasets in Part III. The slight variation observed can be attributed to the low dimensionality of the input data. These findings indicate that MLP demonstrates proficiency in accurately predicting the winding number, affirming its suitability for this task.

### III. REALIZATION OF CHERN STATES IN A CIRCUIT

The Chern Insulator (CI) is a lattice-based system that exhibits topologically protected edge modes, quantized by the lattice Chern number $C$ [43]. This number represents the total charge of Berry flux monopoles and takes integer values. The protection of these topological chiral edge modes relies on the constituent degrees of freedom, the magnitude of the bulk gap, and the prevention of loss.

In this section, we refer to a Chern circuit formed by an electric network, which is shown in Fig. 6 following Ref [45]. The circuit unit cell consists of nodes connected by capacitors and negative impedance converters with current inversion (INICs). The nodes are grounded by inductors and capacitors. The topologically protected chiral voltage edge modes in this circuit resemble a voltage circulator and can be protected from decay by recalibrating gain and loss. Importantly, this circuit structure allows for a topolectrical Chern circuit (TCC) without the need for external bias fields or Floquet engineering. The TCC provides a new approach to realizing topological phenomena in electrical circuits.

The TCC consists of a periodic circuit structure, with each unit cell comprising two nodes connected to adjacent nodes through capacitors and to next-nearest neighbors through INICs. The nodes are grounded by inductors and capacitors. The specific arrangement of circuit components in real space is chosen to preserve connectivity and does not affect any observable quantity. The circuit Laplacian $J_{TCC}(k;\omega)$ connects the vector of voltages measured with respect to ground to the vector of input currents at the nodes of the circuit. In the context of an AC frequency $\omega = 2\pi f$ and a two-dimensional reciprocal space defined by the brick wall gauge, the TCC Laplacian is utilized. In this framework, the expression for $J_{TCC}(k;\omega)$ can be written as:

$$J_{TCC}(k;\omega) = i\omega \left[ (3C_0 + \frac{C_1+C_2}{2} - \frac{1}{\omega^2 L_0})I - C_0[1+\cos(k_x)]\sigma_x - C_0[\sin(k_x)+\sin(k_y)]\sigma_y \right. \\ \left. + \left[ \frac{C_1-C_2}{2} + \frac{2}{\omega R_0}[\sin(k_x)-\sin(k_y)-\sin(k_x-k_y)] \right] \sigma_z \right]. \quad (6)$$

where $C_1 = C_g + \Delta$, $C_2 = C_g - \Delta$, so that $\Delta = \frac{C_1 - C_2}{2}$ can be inversely represented. The Chern number for the lower admittance band is defined as $C = (1/2\pi) \oint d^2 k B(k)$, where $B(k)$ represents the Berry curvature. The theoretical value of C can be calculated by putting $\Delta = \frac{C_1 - C_2}{2}$ into the Chern number calculation formula, which should be [45]:

$$C = \frac{1}{2} \left[ \text{sgn}(\Delta + \frac{3\sqrt{3}}{\omega R_0}) - \text{sgn}(\Delta - \frac{3\sqrt{3}}{\omega R_0}) \right]. \quad (7)$$

Notably, this quantity remains invariant even when the Bravais vector gauge is changed. Similar to the Haldane model, when the two admittance Dirac cones are eliminated by introducing a finite Haldane or Semenoff mass, there is a topologically nontrivial (trivial) region with $C = 1$ ($C = 0$).

We find that C is nonzero when $\omega R_0 < 3\sqrt{3}/\Delta$. In this situation, when positioned within the admittance or eigenfrequency gap and allowing for a boundary termination, a chiral mode emerges. The chiral nature of the voltage boundary mode arises from the fact that $\omega(k) \neq \omega(-k)$, indicating a violation of circuit reciprocity. This violation serves as a necessary condition for $C \neq 0$ in a topolectrical circuit. The introduction of resistances in the circuit breaks time-reversal symmetry, as resistive components experience Joule heating and lead to broken time-reversal symmetry. Circuit reciprocity is defined by the symmetry of the Laplacian in real and reciprocal space. By using operational amplifiers as active circuit elements, the INIC configuration acts as a charge source or sink, breaking reciprocity and introducing an antisymmetric contribution to the Laplacian.

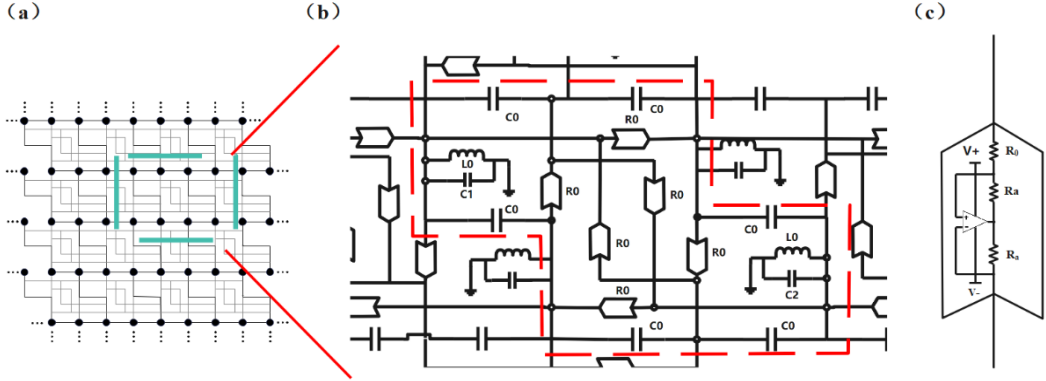

FIG. 6. Topolectrical Chern circuit. (a) The graph representing the alignment of nodes in a brick wall is a three-coordinated circuit. This representation includes both horizontal (*x*) and vertical (*y*) alignments. (b) The detailed structure of the circuit element is described for the rectangular region outlined by the green dashed frame in (a). The circuit element structure is meticulously illustrated within the red dashed framed rectangle. Apart from the capacitive internode connections $C_0$, there exist connections to ground that are both inductive $L_0$ and capacitive $C_{1,2}$. (c) Besides, the structure of the INIC element is op amp. The combination of resistors $R_a$ and $R_0$, along with an operational amplifier with supply voltages $V_+$ and $V_-$, functions as a negative impedance converter with current inversion. In other words, it acts as a positive (negative) resistor from the front (back) end.

In order to more conveniently observe whether the boundary effect occurs in the system, we choose such a Chern circuit system so as to indirectly infer the system topological invariants Chen number *C*. In our work, we build a finite TCC lattice and stimulate with a Gaussian AC current signal centered at $\omega_c = 2\pi f_c$ and with a standard deviation $\Delta\omega_{exc} = 2\pi\Delta f_{exc}$. As shown in Fig. 6(b), The $R_0$, $L_0$, $C_0$, $C_1$, and $C_2$ are indicated as the values of the resistance, inductive, and capacitive, respectively. During data collection, we do not obtain topological features through the Chern number calculation formula but instead through the circuit simulation. We observe the appearance of the voltage of each node of the circuit boundary state, so as to indirectly judge the value of the system term number. Therefore, the data we obtained is essentially experimental in nature. Machine learning algorithms are then applied to explore patterns within our circuit simulation data, rather than relying on theoretical computational methods for system solutions. The theoretical results of the system are only used to compare with the results of the algorithm itself to verify the correctness of the deep learning predictions. The LTSPICE simulations are performed under various configurations with four different injection methods as shown in Fig.7 (a-d), while keeping $\omega_c = \omega_0 = 1/\sqrt{3C_0 L_0}$ constantly. Subsequently, we investigate the influence of circuit components on the overall topological properties of the TCC circuit by adjusting the values of $R_0$, $L_0$, $C_0$, $C_1$, and $C_2$. After manually fine-tuning the simulations, we obtain voltage values of each unit for nearly 1000 different sets of $R_0$, $L_0$, $C_0$, $C_1$, and $C_2$, which constitute our training dataset. This dataset consists of five-dimensional feature inputs, with the circuit structure represented as a matrix of individual units.

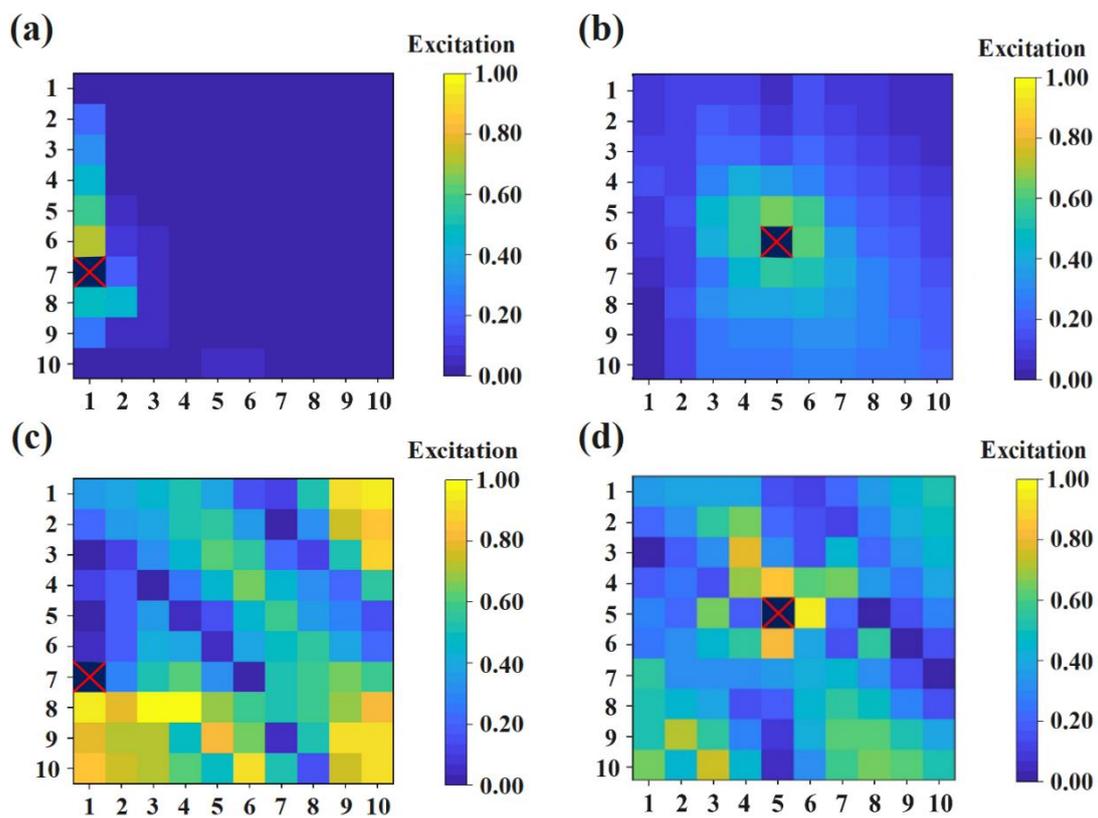

FIG. 7. The simulated voltage pulse in a finite TCC. The position of the current signal within the TCC is emphasized by a red crossed square, as shown in (a)–(d). It is worth noting that all voltage profiles in these figures are normalized relative to the input signal. In (a) - (d), the integrated total voltage signals are analyzed at each cell and simulated using LTSPICE for a (10 × 10) cell grid. (a) $(f_c, \Delta f_{exc}) = (290, 10) kHz$. The integrated total voltage signal is evaluated for a 10×10 unit cell

grid, employing realistic operational amplifiers LT1363, $C_0$=0.1 μF, $L_0$=1 μH, $R_0$=30 Ω, $R_{L_0}$=150 mΩ, $R_{C_0}$=5 mΩ, $C_1$=$C_2$=0 F. (b) With circuit parameters identical to (a), a localized circuit response is observed, but it is important to note that, contrary to (a), injection occurs at the center of the circuit rather than being fed into the boundary modes of the chiral edge. (c) $(f_c, \Delta f_{exc}) = (13.0, 1.0) kHz$. TCC parameters are $C_0$=10 μF, $L_0$=10 μH, $R_0$=10 Ω, $C_1$=$C_2$=0 F. The circuit parameters, denoted as $R_{L_0}$=1 mΩ, $R_{C_0}$=0 mΩ, with the current distributed throughout the entire circuit. (d) With circuit parameters identical to (c), the voltage display exhibits the bulk-boundary correspondence.

Next, we process to construct a neural network for our designed TCC circuit. Due to the enormous size of the training dataset, traditional MLP and CNN models may fail to accurately predict the topological invariants. Therefore, we choose the DenseNet neural network. Unlike MLP and CNN architectures, which are commonly used for one-dimensional and two-dimensional inputs, respectively, the DenseNet network is specifically chosen for its ability to handle the five-dimensional input required by our circuit. This innovation addresses the limitations of MLP and CNN models in accurately predicting the topological invariants of our circuit, which are stored in a 10×10 matrix comprising five distinct features: $R_0$, $L_0$, $C_0$, $C_1$, and $C_2$. Unlike traditional CNN architectures, DenseNet connects each layer in a dense manner, where every layer is directly connected to all previous layers through dense connections. This dense connectivity design endows DenseNet with numerous advantages. For instance, the dense connections enhance the propagation of parameters and gradients, facilitating their efficient flow throughout the network. This design also alleviates the problem of vanishing/exploding gradients, making the network easier to train. In terms of parameter efficiency, DenseNet requires fewer parameters compared to traditional CNN architectures. This is because each layer only needs to learn the residuals relative to the previous layers, rather than learning the entire feature maps. This parameter sharing design allows DenseNet to achieve better performance with the same number of parameters. The design of our DenseNet network and its correspondence with the circuit are shown in Fig. 8. This choice represents a significant innovation in our research.

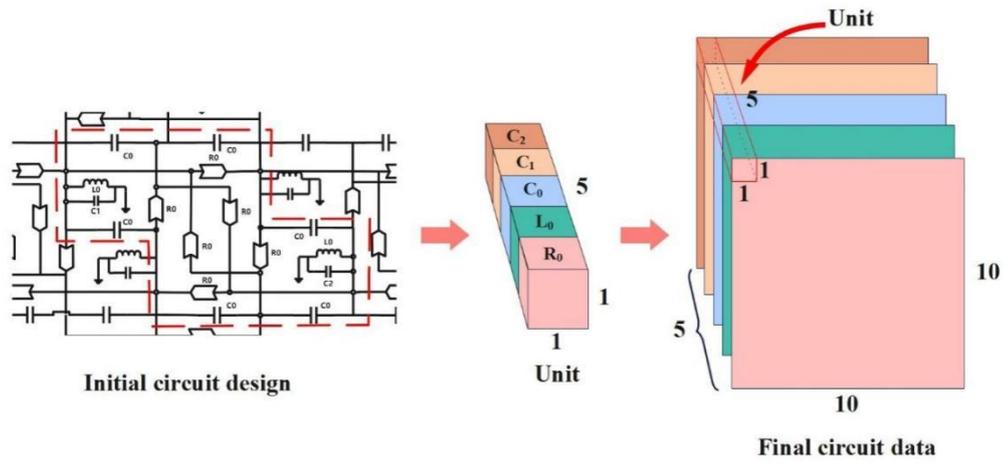

FIG. 8. The constituents and sequence of the final circuit data, wherein each unit of the corresponding circuit is represented by a matrix. This matrix stores five distinct features, namely $R_0$, $L_0$, $C_0$, $C_1$, and $C_2$. Additionally, each sample comprises a 10×10 matrix with a dimensionality of five.

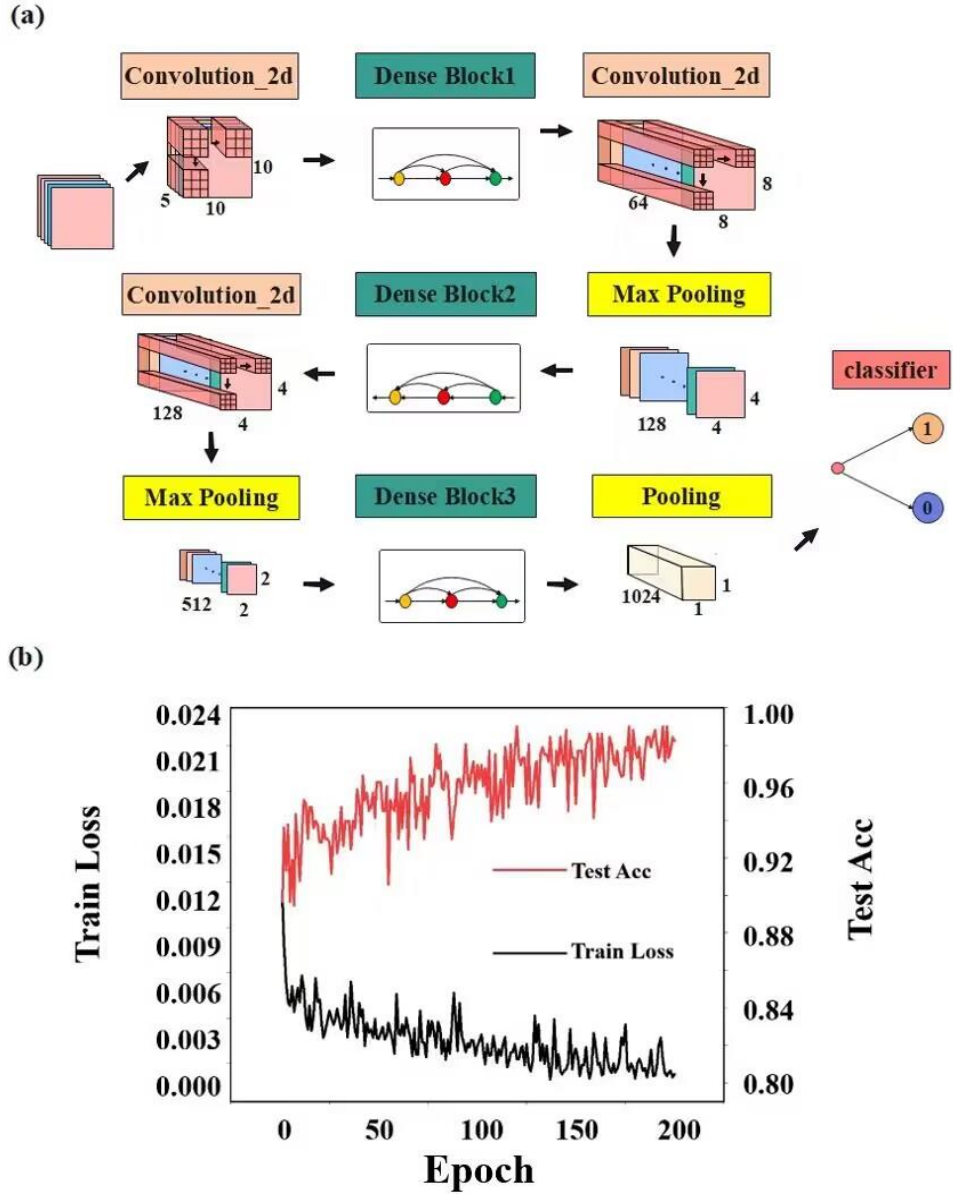

FIG. 9. DenseNet for 5D feature input of the TCC. (a) A DenseNet neural network with three Conv2D layers, two dense block layers and three pooling layers is constructed. (b) The train loss and test acc of each epoch of the DenseNet network.

As shown in Fig. 9(a), the network begins with a convolutional layer, which employ a 3×3 kernel, to process a dataset of size 10×10×5. This initial convolutional layer aims to extract important features from the input data. Following the convolutional layer, a dense block is employed. The dense block is a type of layer that promotes feature reuse and boosts information flow throughout the network. It enhances the network's representational power by concatenating the outputs of multiple convolutional layers, allowing for more complex feature extraction. After the dense block, another convolutional layer is applied using a 3×3 kernel on a dataset of size 8×8×64. This convolutional layer further refines the extracted features. Subsequently, a max pooling layer is employed to reduce the spatial dimensions of the data. This pooling layer reduces the size of the data to 128 feature maps, each with a size of 4×4. Another dense block follows the pooling layer, further enhancing the network's ability to capture intricate patterns and features. Next, another convolutional layer is applied using a 3×3 kernel. This convolutional layer continues to extract more refined features from the data. Continuing with the network structure, a final dense block is employed, further enriching the network's representation capabilities. Finally, a pooling layer is used to flatten the data, converting it into a suitable format for classification tasks. Upon sufficient training, it becomes evident that the accuracy of this network consistently remains at approximately 99% as depicted in Fig. 9(b). Furthermore, the convergence of the loss during the training process is remarkably rapid, indicating a favorable training outcome and a successful establishment of the model.

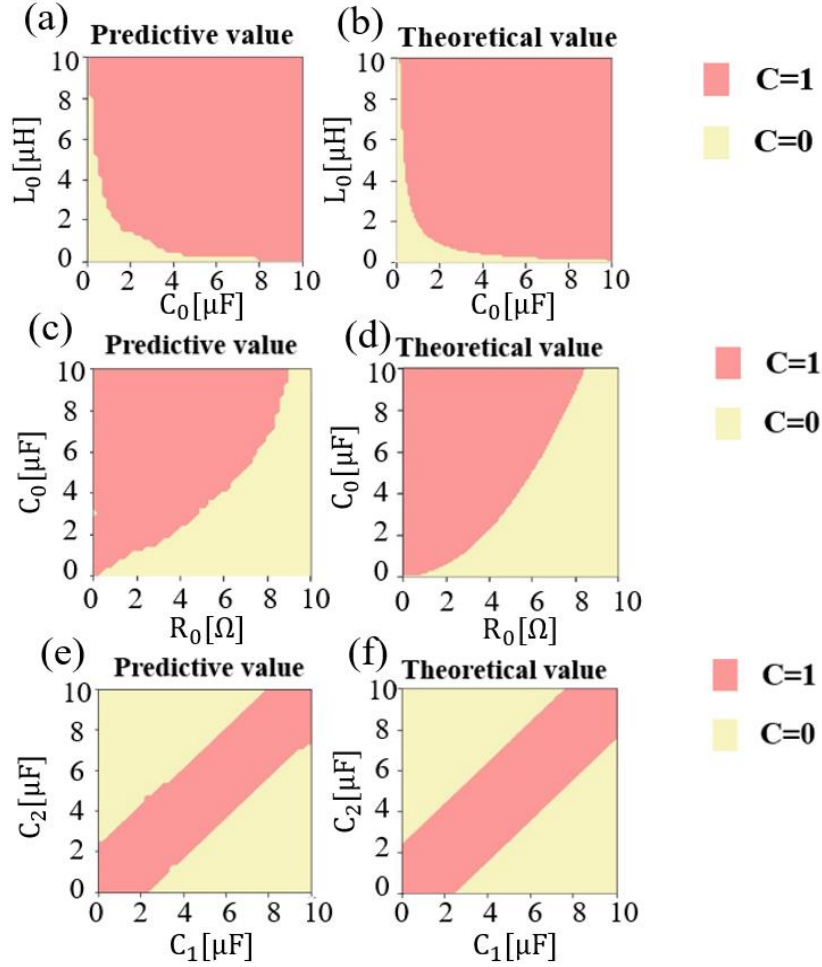

FIG. 10. The predicted phase diagrams of the DenseNet. Due to the limitation of representing five-dimensional data in a visual graph, we choose three sets of parameters to construct predicted phase diagrams and compare them with the theoretical values calculated via Eq. (6). For (a) and (b), we set $R_0$=5 $\Omega$, $C_1$=6 μF, $C_2$=1 μF. For (c) and (d), the parameters are $L_0$=0.2 μH, $C_1$=1 μF and $C_2$=4 μF. Lastly, for (e) and (f), we set $C_0$=0.8 μF, $L_0$=0.8 μH, and $R_0$=6 $\Omega$.

Our results are presented in Fig. 10, which show a comparison between the calculated phase diagrams (panels a, c, d) and predicted phase diagrams (panels b, e, f). The pink and yellow in the phase diagrams correspond to Chern numbers of 1, and 0, respectively. There are only a few tiny patches of incorrect predictions, which occur predominantly near the phase boundaries between regions with different Chern numbers. It proves that our neural network is reliable and suitable for predictions involving higher dimensions.

Subsequently, to explore the reliability and generalization of our model, we randomly select a set of circuit data and remove a 2×2 unit block at the bottom of the circuit (as shown in Fig. 11). For the network, this corresponds to missing matrix data at the corresponding position of the original circuit. Using the trained network, we predict the state of the defective circuit. The circuit data is the same as in Fig. 7(a). The network predicted a state value of $C=1$, indicating that the voltage should theoretically exhibit a boundary state distribution.

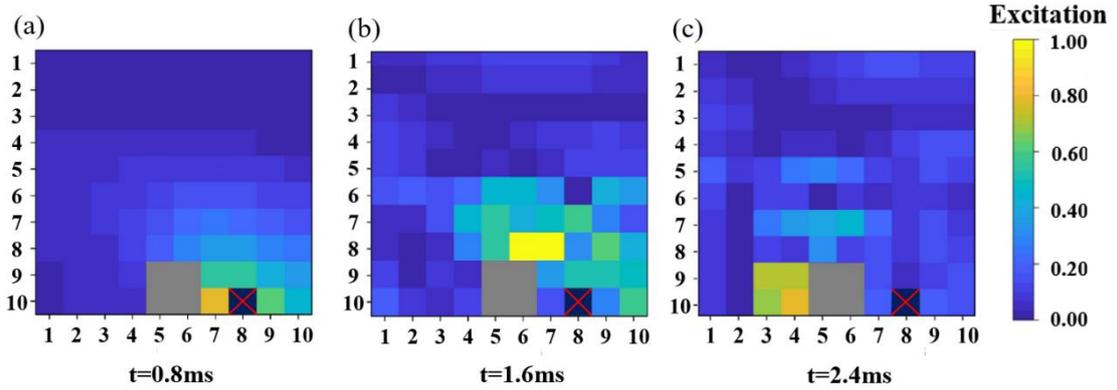

FIG. 11. (a-c) The heat maps of the voltage values for each unit in the TCC circuits with defective states at t=0.8 ms, 1.6 ms, and 2.4 ms, respectively. The circuit parameters of the TCC are identical to those illustrated in FIG. 7(a), while the defect area, represented by the gray region, is implemented by grounding the corresponding nodes.

Additionally, this successful prediction of the state value for circuits with defects further confirms the robustness and versatility of our network. It indicates that our network is capable of comprehensively exploring internal relationships within the data and demonstrates its accuracy in handling complex scenarios even in the presence of missing or inaccurate data. This capability holds great significance in practical applications, as it allows for the effective analysis and diagnosis of circuit behavior, enabling engineers and researchers to make informed decisions and optimize circuit designs.

## IV. SUMMARY AND OUTLOOK

In summary, we successfully train neural networks that learns topological invariants from the large data set of Hamiltonians in the momentum space. For the one-dimensional non-Hermitian SSH model, we train MLP and CNN neural network to predict the different phases characterized by their winding numbers. Each neural network owns an accuracy greater than 98%. For the two-dimensional non-Hermitian Chern lattice model, we construct and train a DenseNet network to yield excellent accuracy in predictions of the topological invariants, the Chern number. The DenseNet network, with its dense connectivity and ability to handle the five-dimensional input required by our circuit, exhibits an impressive accuracy of 99.5%. This achievement highlights a significant innovation and departure from traditional approaches. Based on the predicted result, a topolectrical Chern circuit is built to confirm the nonreciprocity and the topological protection characteristic. It is also possible to generalize our proposed methods to the non-Hermitian topological model in higher dimension [48-49].

Before concluding, we want to make some remarks on the role of applying deep learning to topological phases. In almost all the previous works [5, 27-42], the deep learning methods usually are generalized to predict the different topological phases. The connection between topological invariants (the winding number for one-dimension model, and the Chern number for two-dimension model) and their corresponding topological models (SSH Model and Chern circuit system) has been proven to be deducible through rigorous mathematical formulas [12,45]. Thus, we can consider it as the ground truth. Similarly, the topological characteristics such as the skin effect [35], nonreciprocity, and topological protection [45,50], can also be chosen as the ground truth due to their well-established mathematical theories. It is clear that ground truth must involve rigorous mathematical deductions. In other words, transforming an experimental phenomenon into a ground truth requires rigorous mathematical derivations, although this is often complex and challenging. Pleasantly enough, deep learning can assist scientists in uncovering regularities from experimental phenomena in topology, which can provide insights for establishing rigorous mathematical models [51].

**Acknowledgements**
The authors thank for the support by National Natural Science Foundation of China under (Grant 62001289), NUPTSF (Grants No. NY220119, NY221055), Jiangsu Provincial Double-Innovation Doctor Program (Grants No. JSSCBS20210546).